\def\year{2015}
\documentclass[letterpaper]{article}
\usepackage{aaai}
\usepackage{times}
\usepackage{helvet}
\usepackage{courier}
\usepackage{grffile}	
\usepackage{graphicx}
\usepackage{epstopdf}
\usepackage[table]{xcolor}
\usepackage[hyphens]{url}
\usepackage{subfig}
\usepackage{pifont}
\usepackage{multirow}

\DeclareTextFontCommand{\textbfit}{%
  \fontseries\bfdefault 
  \itshape
}

\title{Characterizing Information Diets of Social Media Users}

\author{Juhi Kulshrestha,~ Muhammad Bilal Zafar,~ Lisette Espin Noboa,\\{\bf \Large Krishna P. Gummadi,} ~ {\bf \Large Saptarshi Ghosh}\\
Max Planck Institute for Software Systems\\
Kaiserslautern-Saarbruecken, Germany
}

\begin{document}

\maketitle

\begin{abstract}
\begin{quote}
\noindent 
With the widespread adoption of social media sites like Twitter and Facebook, there has been a shift in the way
information is produced and consumed.
Earlier, the only producers of information were traditional news organizations, which
broadcast the same carefully-edited information to all consumers over mass media channels. 
Whereas, now, in online social media, any user can be a producer of information, 
and every user selects which other users she connects to, thereby choosing the information she consumes. 
Moreover, the personalized recommendations that most social media sites provide 
also contribute towards the information consumed by individual users.
In this work, we define a concept of {\it information diet} -- which is the topical distribution of a given
set of information items (e.g., tweets) --
to characterize the information produced and consumed by various types of users in the popular Twitter social media.
At a high level, we find that 
(i)~popular users mostly produce very specialized diets focusing on only a few topics; in fact, 
news organizations (e.g., NYTimes) produce much more focused diets on social media
as compared to their mass media diets,
(ii)~most users' consumption diets are primarily focused towards one or two topics of their interest, and
(iii)~the personalized recommendations provided by Twitter help to mitigate some of the topical
imbalances in the users' consumption diets, by adding information on diverse topics apart from the users' primary topics of interest.
\end{quote}
\end{abstract}

\section{Introduction} \label{sec:intro}

The rapid adoption of social media sites like Twitter and Facebook is
bringing profound changes in the ways information is produced and
consumed in our society.  Traditionally, people acquired information
about world events via {\it mass media}, i.e., dedicated news
organisations that relied on some {\it broadcast} medium like print
(NYTimes or Economist), radio (NPR, BBC radio), or television (CNN,
ESPN) to disseminate the information to large numbers of users.  
Mass media communications are characterised by (i)~a small number (few tens
to a few hundreds) of news organisations controlling what hundreds of
millions of users consume, (ii)~an expert team of editors at each news
organisation carefully vetting and selecting news stories to ensure a
{\it balanced coverage} of important news stories, and (iii)~all
consumers receiving the same standardised information broadcast by
each mass media source.

In contrast to the organised world of information production and
consumption in broadcast mass media, online social media sites like
Twitter and Facebook offer a chaotic information marketplace for
millions of producers and consumers of information. Unlike mass media,
in social media (i)~any of the hundreds of millions of users of these
systems can be a producer as well as a consumer of information, 
(ii)~these individual users are {\it not} expected to provide a balanced
coverage of news-stories -- they publish any information that they
deem important or necessary to share with their friends in real-time, and
(iii)~information consumption is {\it personalised} and not all users
consume the same information -- every individual user selects (e.g.,
by establishing social links) her preferred sources of information
from the millions of individual producers, and {\it recommender
  systems} deployed by social media platforms provide an additional
source of information to the user.  Thus, individual social media
users might receive information that is not only unbalanced in terms
of coverage of news-stories, but is also very different from what
other users in the system receive.

An entire discipline, {\it media studies}, has largely focused on
analysing the coverage of information published on broadcast mass
media and how it impacts the consumers of mass media. 
In contrast, research on understanding the composition of
information produced and consumed by social media users is still in
its infancy, being limited to a few macroscopic studies on the
amounts of information posted by broad categories of users (e.g., celebrities)~\cite{Wu_whoSaysWhatTwitter,Kwak_www10}.
There has not been much work on analysing the {\it composition}
of the information produced or consumed by users at the granularity of individual messages.

In this paper, we take the first step towards addressing this
challenge by defining the notion of {\it information diet}. 
Similar to diet in nutrition, information diet of a user
refers to the composition of all the information consumed or
produced by the user~\cite{info-diet-book}. Specifically, we focus on the {\it topical
  composition} of users' diets, i.e., the fraction of their
information diets that correspond to different topical categories of
information (e.g., information on politics, sports, entertainment, and
so on).

One of our key goals is to better understand how the differences in
information production and consumption processes between broadcast
mass media and online social media affect users' diets. So we
conducted a comparative analysis of the topical compositions of the
information diets produced, consumed, and recommended on social media
and the mass media.  Our investigation focused on the following three
high-level questions:

\begin{enumerate}

\item {\bf Production}: What is the topical composition of information
  published on broadcast mass media (e.g., NYTimes print edition)? How
  does the information produced by social media accounts compare with
  the information published on mass media?

\item {\bf Consumption}: How balanced or unbalanced are consumption
  diets of social media users (relative to mass media diet)? Are
  users' consumption diets heavily skewed towards a few topics of
  their interest, or do they also tend also receive information on a broad
  variety of topics covered in mass media?

\item {\bf Recommendations}: Do personalised recommender systems
  deployed by the social media platform provide balanced or unbalanced diets
  (relative to mass media) to social media users? Do they mitigate or
  exacerbate the imbalances in the users' consumption diets?
    
\end{enumerate}

\noindent We attempt to address the above questions in the context of the
Twitter social media platform. To conduct our study, we needed a
methodology to infer the topics of individual posts on Twitter.  The
bounded length of tweets makes it challenging to infer topics at the
level of individual tweets.
We propose a novel methodology to infer the topic of a post by 
leveraging the topical expertise of the Twitter {\it users} who have posted it. 
To obtain the information about users'
topical expertise, we leverage a methodology based on Twittre Lists,
developed in our prior works~\cite{cognos_sigir,whoiswho_wosn}.
We show that our methodology performs better at inferring topics for
posts than a state-of-the-art publicly deployed commercial topic
inference system.

Our study conducted using our above methodology yields several key
insights. We highlight a few below:

\begin{enumerate}

\item Mass media sources cover a wide range of topics from {\it
  politics} and {\it business} to {\it entertainment} and {\it
  health}. But on social media, the individual sources of information
  are very focused and publish information dominated by a few
  topics. It is up to the social media users to select sources to
  obtain a balanced diet for themselves.

\item We find that for most users, a large fraction of their consumed
  diet comes from as few as one or two topics, and they hear very little
  about other niche topics like {\it health} and {\it environment} (unless they are
  interested in these topics).

\item We find that social recommendations, i.e., recommendations about
  information popular in a user's social network neighbourhood~\cite{twitter-reco}, often
  do not match the user's preferred diet. The differences between
  recommended and consumed diets are likely due to differences in
  the interests of a user and the interests of her network neighbours. As a result,
  social recommendations introduce topical diversity to a user's diet
  and can help balance its topical composition.

\end{enumerate}

\noindent 
We have publicly deployed a Web-based service for measuring the information diets produced and consumed by Twitter users, at {\it http://twitter-app.mpi-sws.org/information-diets/}.

Our work and findings have a number of important implications. 
As social media becomes more popular, it is important to raise
awareness about the balance or imbalance in information diets produced
and consumed on social media. Our findings raise the need for better
information curators (human editors or automated recommendation
systems) on social media that provide a more balanced information
diet.
Finally, our work is an early attempt, and much future
work still remains to be done both on understanding the impact of the
diets on consumers in shaping their opinions and on other ways for
quantifying the diets beyond topical composition.

\section{Related Work} \label{sec:related}

\noindent{\bf Analysis of content on mass media:}
Media studies has been an active
field which analyzes the content coverage on mass media, and its effects 
on the society.\footnote{{\it http://en.wikipedia.org/wiki/Media\_studies}}  
There exist a number of `media watchdog organizations' (e.g., FAIR ({\it http://fair.org/}), AIM ({\it http://www.aim.org/})) 
which judge the content covered by news organizations based on fairness, balance and accuracy.
Additionally, there have also been studies on media biases~\cite{groseclose2005measure,budak2014fair}. 
Such studies are easier to perform over mass media since it is a broadcast medium and all users receive the same information.
On the other hand, studying the information consumed on social media is much more challenging since individual users shape their own personalized channels of information by selecting the other users to follow.

~\\
\noindent{\bf Information production \& consumption on social media:}
Prior studies on information production and consumption
on social media~\cite{Wu_whoSaysWhatTwitter,Kwak_www10,cha@cybernets2012} 
have been limited to studying the amount of information being exchanged among various users. 
There has not been any notable effort towards analyzing the {\it topical composition} of the information 
produced or consumed, which is the goal of this work.

There have also been some prior works on whether social media users are receiving multiple perspectives on a specific 
event or topic~\cite{Balasubramanyan:icwsm12,Conover:icwsm11,park2009newscube,Adamic:LinkKDD:2005,weber_egypt_twitter}.
Though we focus only on the topical composition of the information produced and consumed
by social media users, 
the concept of information diet introduced in this work can be extended to 
study opinion polarization on social media.

~\\
\noindent{\bf Topic inference of social media posts:}
To our knowledge, all prior attempts to infer the topic of a tweet / hashtag / trending topic
rely on the content itself -- 
either applying NLP and ML techniques~\cite{quercia2012tweetlda,Ramage_TweetTopicModel,ottoni2014pins,zubiaga2011classifying} 
or mapping to external sources such as Wikipedia or Web search results~\cite{meij_TweetSemantics,bernstein_eddi} 
-- in order to infer the topics. 
Such methodologies are of limited utility in the case of social media like Twitter,
primarily due to the tweets being too short, and the informal nature
of the language used by most users~\cite{whoiswho_wosn,wagner_topic_expertise}.
In contrast to these previous approaches which focus on the content, 
our methodology focuses on the characteristics of the {\it authors} of the content to infer its topic.

\section{Methodology: Quantifying Information Diets} \label{sec:methodology}

In this paper, we introduce the notion of {\it information diet} of a set of 
information items (e.g. a set of tweets or hashtags), as the topical composition of the information items.
We define the topical composition over a given set of topics
as the fraction of information related to each topic.
In this section, we present our methodology for quantifying the information diet for a set of tweets on Twitter.

We chose {\it hashtags} and {\it URLs} as the basic elements of information in a tweet and collectively refer to them as {\it keywords}. 
However, our methodology can be easily extended to include other kinds of keywords such as named entities. 
To justify our choice of keywords, we conducted a survey through 
Amazon Mechanical Turk (AMT: {\it https://www.mturk.com/}), 
where we showed workers 500 randomly selected tweets from Twitter's 1\% random sample which did {\it not} contain any keyword. 
A majority of the AMT workers judged 96\% of the tweets without any keywords to be non-topical, i.e., 
they mostly contained conversational babble. 
Thus, the hashtags and URLs contain crucial signals about the topicality of tweets, justifying our 
decision to only consider hashtags and URLs as keywords for inferring the topic of tweets. 

The key step in our methodology for quantifying information diets consists of 
inferring the topic of a keyword, which is described next.

\subsection{Inferring topic of a keyword} 

As discussed in the Related Work section, prior approaches for inferring 
the topic of a tweet / keyword rely on the content itself.
Such approaches tend to perform poorly on short posts containing informal language~\cite{whoiswho_wosn,wagner_topic_expertise}.
So we propose a different technique to infer the topic of a keyword 
which relies on the topical expertise of the users who are discussing that keyword.
The basic intuition behind our technique is that if
many users interested in a certain topic
are discussing a particular keyword, that keyword is most likely related to that topic.

To identify the topical expertise of users in Twitter,
we leveraged the List-based methodology developed in our prior works~\cite{whoiswho_wosn,cognos_sigir}
to retrieve {\it expertise tags} for topical experts. 
For instance, some of the tags inferred by this methodology for the expert 
@ladygaga are `music', `entertainment', `singers', `celebs' and `artists'. 
We extracted topical expertise of 771,000 experts on Twitter by using this methodology.
The details of the methodology are omitted here for brevity.

\begin{table}
\center
\small
\begin{tabular}{|l|l|l|}
\hline
{\bf Topic categories} & \multicolumn{2}{c|}{{\bf Some related terms}} \\
\hline
Arts-crafts  &  \multicolumn{2}{l|}{art, history, geography, theater, crafts, design} \\
\hline
Automotive  &  \multicolumn{2}{l|}{vehicles, motorsports, bikes, cars} \\
\hline
Business-finance  & \multicolumn{2}{l|}{retail, real-estate, marketing, economics}  \\
\hline
Career  & \multicolumn{2}{l|}{jobs, entrepreneurship, human-resource}  \\
\hline
Education-books  & \multicolumn{2}{l|}{books, libraries, teachers, school}  \\
\hline
Entertainment  &  \multicolumn{2}{l|}{music, movies, tv, radio, comedy, adult} \\
\hline
Environment  &  \multicolumn{2}{l|}{climate, energy, disasters, animals} \\
\hline
Fashion-style  &  \multicolumn{2}{l|}{style, models} \\
\hline
Food-drink  &  \multicolumn{2}{l|}{food, wine, beer, restaurants, vegan}  \\
\hline
Health-fitness  &  \multicolumn{2}{l|}{disease, mental-health, healthcare} \\
\hline
Hobbies  &  \multicolumn{2}{l|}{photography, tourism, gardening} \\
\hline
Paranormal  & \multicolumn{2}{l|}{astrology, supernatural}  \\
\hline
Politics-law  &  \multicolumn{2}{l|}{politics, law, military, activism} \\
\hline
Religion  &  \multicolumn{2}{l|}{christianity, islam, hinduism, spiritualism} \\
\hline
Science  & \multicolumn{2}{l|}{physics, chemistry, biology, mathematics}  \\
\hline
Society  & \multicolumn{2}{l|}{charity, LGBT}  \\
\hline
Sports  &  \multicolumn{2}{l|}{football, baseball, basketball, cricket} \\
\hline
Technology  &  \multicolumn{2}{l|}{mobile-devices, programming, web-systems} \\
\hline
\end{tabular}
\caption{{\bf The 18 topic-categories to which keywords / tweets will be mapped, and some 
terms related to each topic. The terms will be matched with expertise-tags.}} 
\label{tab:topic-categories}
\end{table}

Next, we used two standard topical hierarchies --
the Open Directory Project ({\it www.dmoz.org}) and AlchemyAPI ({\it www.alchemyapi.com/api/taxonomy/}) --
 to obtain 18 topical categories and their related terms, as shown in
 Table~\ref{tab:topic-categories}.
The 18 topical categories were selected by combining the top categories of the 
 two hierarchies,  
 while the related terms were derived from their lower levels.
 In the rest of the paper, we quantify information
 diets by inferring the fraction of information from each of these 18 topics.
We also mapped the experts to one or more of the 18 topic
 categories, by matching the inferred tags of each expert to the related terms of the topical categories.

As stated earlier, the main intuition behind our methodology is that
if several experts on a topic are posting a keyword, then that keyword is
most likely related to that topic. 
To infer the topic of a keyword $k$,
we first identify the set of experts $E_k$ who have posted $k$.
We do {\it not} attempt to infer the topic of a keyword unless it has been posted by at 
least 10 of our identified experts.
For each topic $t$ (in Table~\ref{tab:topic-categories}), 
we then determine the fraction ($f_t$) of experts in $E_k$ who are mapped to that topic $t$.
Next, to account for the varying number of experts mapped to different topics,
we normalize the fraction $f_t$ by the total number of experts on topic $t$ in our data set.
Finally, we select the topic with the highest normalized fraction $f_t$
to be the inferred topic of keyword $k$.
Further details of the methodology can be found at 
{\it http://twitter-app.mpi-sws.org/information-diets/}.

\subsection{Evaluating the topic inference methodology}  \label{sub:eval}

We now present the evaluation of the performance of our 
proposed topic inference methodology, and compare its performance
with that of a state-of-the-art commercial service, AlchemyAPI, that uses NLP and deep-learning techniques for topic inference.
We found the performances to be very similar
for both hashtags and URLs; hence, for brevity, 
we only present the evaluation results for hashtags.

The set of hashtags used for evaluation is derived from the Twitter 1\% random sample~\footnote{We considered only English tweets, i.e., tweets in which
at least half of the words occur in a standard English dictionary.}
from a week in December 2014. It consists of: (i)~200 popular hashtags which were most tweeted, and (ii)~200 randomly selected hashtags.
We inferred the topic of a hashtag using AlchemyAPI by passing 1000 randomly selected tweets
containing the hashtag. 
Table~\ref{tab:proposed-alchemy-comparison-htags} compares the 
performance of the proposed methodology with AlchemyAPI, based on two metrics - coverage and accuracy.

\begin{table}
\center
\small
\begin{tabular}{|l |l |l |l |}
\hline
{\bf Metric} & {\bf Methodology}		& \multicolumn{2}{c|}{\bf Hashtags} \\ \cline{3-4}
	 	&	& {{\bf Popular}} & {\bf Random}  \\
\hline
\hline
\multirow{2}{*}{\bf Coverage} & AlchemyAPI   & 22.5\%  & 55.5\%   \\ \cline{2-4}
 & Proposed	& 98\%	  & 82.5\%  	\\
\hline
\multirow{2}{*}{\bf Accuracy} & AlchemyAPI	& 44.44\% &  51.35\% \\ \cline{2-4}
 & Proposed	& 58.67\% &  49.69\%  \\
\hline
\end{tabular}
\caption{{\bf Comparing the proposed topic inference methodology with AlchemyAPI (which uses NLP techniques) in terms of
coverage and accuracy.}}
\label{tab:proposed-alchemy-comparison-htags}
\end{table}

\noindent {\bf Coverage:} It is defined as the fraction of keywords for which a methodology is able to infer a topic. 
Table~\ref{tab:proposed-alchemy-comparison-htags} shows that
our proposed methodology performs significantly better than AlchemyAPI, which 
possibly fails due to the informal and abbreviated language used in most tweets.
Note that our methodology is able to infer topics for a relatively smaller fraction of 
random hashtags than the popular ones, since we need the hashtag to be posted by at least 10 experts.

\noindent {\bf Accuracy:} It is defined as fraction of keywords for which the inferred topic is relevant.
Relevance was judged through an AMT survey --
we showed the hashtag, 20 random tweets containing the hashtag, and the inferred topic to 
five AMT workers and asked them to judge if the inferred topic of the hashtag is relevant. 
Table~\ref{tab:proposed-alchemy-comparison-htags} shows the majority opinion of the five workers --
the proposed methodology is accurate for a larger fraction of popular hashtags, while
AlchemyAPI performs slightly better for randomly selected hashtags.

Overall, our proposed methodology performs
better than a state-of-the-art NLP-based technique in inferring topics of hashtags,
especially for popular ones --
not only does the proposed methodology infer topics for more hashtags,
but also the inferred topics are more accurate.

\subsection{Quantifying information diet of social media posts}  \label{sub:infodiet-quantify}

Having established the methodology to infer the topic of a keyword, we now use it to
construct the information diet of a set of tweets. We first extract the
keywords from every tweet in the set and infer the topic of each individual keyword.
We then construct a {\it topic-vector} for the given set of tweets, where the 
weight of a topic is the total contribution of all keywords inferred to be on that topic.
Since a tweet can contain multiple keywords, we 
normalize the contribution of each keyword within a tweet by the number of keywords in that tweet
(so that each tweet contributes a total weight of 1 to the topic-vector).
This topic-vector represents the information diet of the given set of tweets.

\subsection{Limitations of our methodology}
We briefly discuss some limitations in our approach of quantifying the information diets of users.
First, since we infer the topics of only those keywords which have
been tweeted by at least 10 topical experts, we
have a lower coverage and accuracy for non-popular keywords.
However, the later sections show that the popular information forms a large 
fraction of users' diets; hence, the approach is likely to be able to estimate
the information diets of users fairly accurately.

Second, while we only focus on information that a user posts or consumes on
Twitter, we are aware that a user in Twitter is also likely to get
information from other online as well as off-line sources.
However, as users are relying more and more on social
media sites such as Twitter and Facebook to find interesting information~\cite{media-consumption-thru-social-media},
what a user consumes in Twitter is likely to be an increasingly significant factor in shaping her overall information diet.

\section{Mass Media Diet}

\noindent As mentioned earlier, the goal of this study is to compare and contrast 
the processes of production and consumption 
of information over broadcast mass media and over social media. 
We analyze the information being published over mass media by three popular news organizations 
-- NYTimes, Washington Post and The Economist.
We collected their broadcast print editions for three days in December 2014, and categorized the news-articles
into our 18 topic-categories (Table~\ref{tab:topic-categories}) through human feedback.
Each news-article was shown to five distinct workers recruited through AMT, and the majority verdict
was considered as the topic for the news-article.

\begin{table}
\small
\center
\begin{tabular}{|l|c|c|c|}
\hline	 
{\bf Topic}	 & {\bf NYTimes} & {\bf Wash. Post} & {\bf Economist}  \\ 
\hline 
Arts-Crafts & 4.56\% & 0.0\% & 1.85\%  \\ 
\hline 
Automotive & 1.34\% & 0.0\% & 0.37\%  \\ 
\hline 
\textbfit{Business-Finance} & \cellcolor{red!20}7.51\% & \cellcolor{red!20}8.65\% & \cellcolor{red!20}28.04\%  \\ 
\hline 
Career & 0.8\% & 0.48\% & 0.74\%  \\ 
\hline 
Education-Books & 1.88\% & 5.29\% & 3.32\%  \\ 
\hline 
\textbfit{Entertainment} & \cellcolor{red!20}12.33\% & \cellcolor{red!20}13.94\% & 1.48\%  \\ 
\hline 
Environment & 3.49\% & 0.96\% & 7.01\%  \\ 
\hline 
Fashion-Style & 0.0\% & 1.44\% & 0.0\%  \\ 
\hline 
Food-Drink & 4.83\% & 6.25\% & 2.21\%  \\ 
\hline 
Health-Fitness & 6.17\% & 5.29\% & 2.95\%  \\ 
\hline 
Hobbies-Tourism & 1.34\% & 0.0\% & 0.37\%  \\ 
\hline 
Paranormal & 0.27\% & 0.0\% & 0.0\%  \\ 
\hline 
\textbfit{Politics-Law} & \cellcolor{red!20}29.49\% & \cellcolor{red!20}37.5\% & \cellcolor{red!20}35.06\%  \\ 
\hline 
Religion & 2.14\% & 0.96\% & 2.95\%  \\ 
\hline 
Science & 1.34\% & 0.96\% & 2.58\%  \\ 
\hline 
Society & 3.75\% & 6.73\% & 3.32\%  \\ 
\hline 
\textbfit{Sports} & \cellcolor{red!20}15.01\% & \cellcolor{red!20}9.62\% & 1.11\%  \\ 
\hline 
Technology & 3.75\% & 1.92\% & 6.64\%  \\ 
\hline 
\end{tabular}
\caption{{\bf Mass media information diets of three news organizations, where the topics of the news-articles were judged by AMT workers (top topics highlighted).}}
\label{table:mass-media-diet}
\end{table}

Table~\ref{table:mass-media-diet} shows the mass media information diets of the three news organizations. 
We find that all the news organizations tend to focus 
(i.e., post majority of their news-articles) 
on a few popular topics -- politics, entertainment, and sports for NYTimes and Washington Post, and mainly politics and business-finance for The Economist. 
However, despite their bias towards these few popular topics, the mass media diets also 
have a spread over the remaining less popular topics --
the 12 least popular topics contribute 25\% of the diet for NYTimes and 17\% for both Washington Post and Economist. 

In the following sections, we use these mass media information diets as a baseline 
for comparing with various information diets on social media.

\section{Production: Social vs. Mass Media Diets}  \label{sec:mass media}

Traditionally, in mass media, editors of news-organizations are expected to 
ensure that the news-stream has a balanced coverage across various topics of interest of the subscribers,
by following definite guidelines.
In contrast, every user-account in social media serves as a producer / source of information,
and there are no definite guidelines on the content being posted by any account.
To analyze the effects of these differences, this section compares various information diets
being produced in social media with those of mass media (described in the previous section).

\begin{figure*}[tb]
\center{
\subfloat[{\bf NYTimes }]{\includegraphics[height=0.32\textwidth,width=3.5cm,angle=-90]{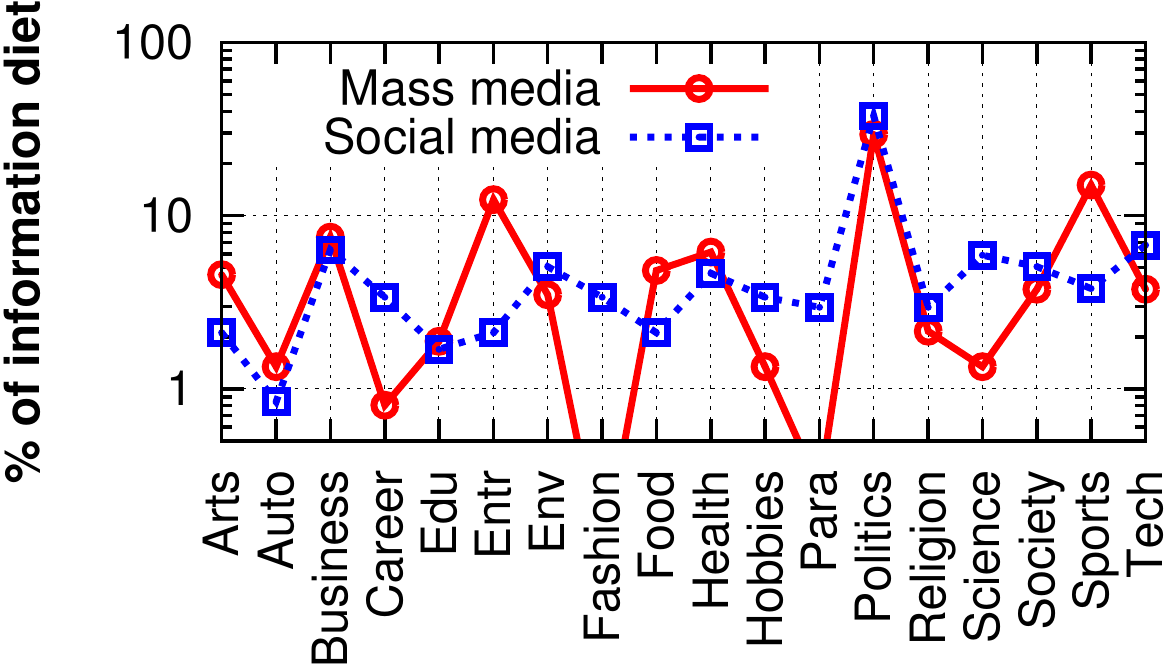}}
\hfil
\subfloat[{\bf Washington Post}]{\includegraphics[height=0.32\textwidth,width=3.5cm,angle=-90]{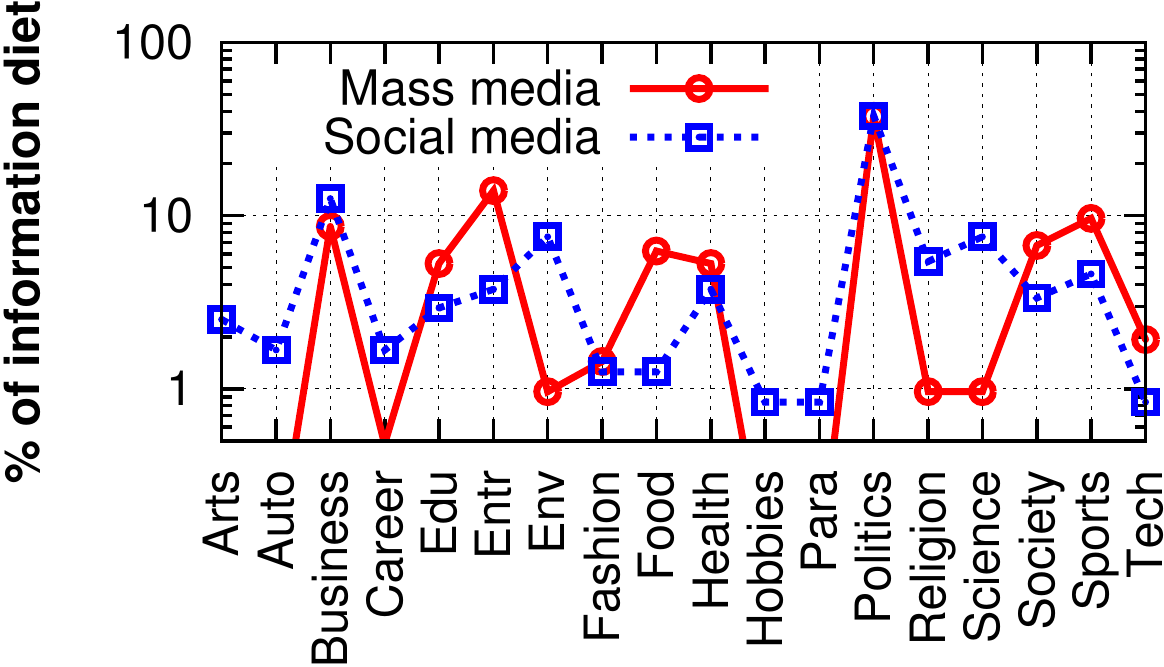}}
\hfil
\subfloat[{\bf Economist}]{\includegraphics[height=0.32\textwidth,width=3.5cm,angle=-90]{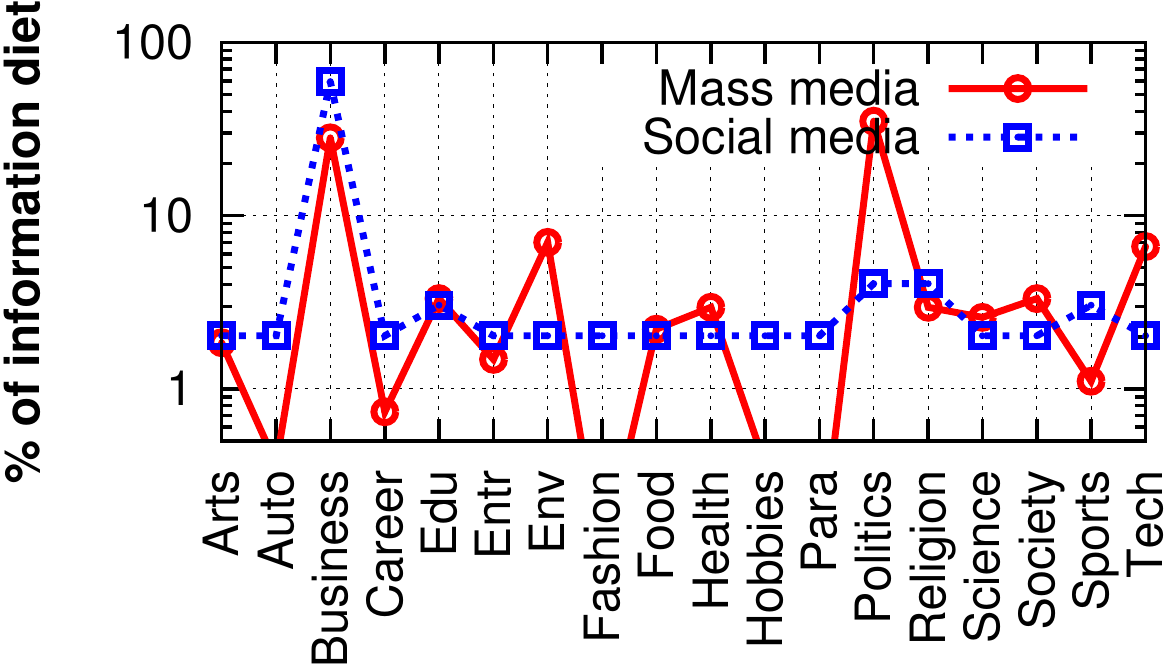}}
}
\caption{{\bf Comparing the information diet posted by news organizations in mass media (news-articles in print editions) and social media (tweets posted by the primary Twitter accounts) for the same days in December 2014. Topics with contribution less than 0.5\% not shown.}}
\label{fig:media-compare}
\end{figure*}

\subsection{News organizations: Social media vs. mass media}

We first address the question: {\it are there differences between the information diets published by news organizations over mass media and social media?}
To answer this question, we collected the tweets posted by the 
Twitter accounts of the three selected news organizations (NYTimes, Washington Post and The Economist) during December 2014, 
and generated the information diet produced by these news organizations over social media.\footnote{The statistics
presented in this section are for the same three days in December 2014, over which the mass media diets were analyzed
in the previous section. However, we observed that the information diets remain relatively unchanged over longer time-durations.}

\begin{table}
\center
\small
\begin{tabular}{|l |l |l |l |}
\hline
{\bf Social media} & {\bf Topic of} 	  & \multicolumn{2}{c|}{\bf Contribution of topic} \\ \cline{3-4}
{\bf account}	   & {\bf specialization} & {\bf Social}  & {\bf Mass} \\
		   & 			  & {\bf media}   & {\bf Media} \\
\hline
NYTSports & Sports & 66.6\% & 	15.0\%\\
\hline
 nytimesbusiness   & Business & 66.1\%	& 7.5\% \\
\hline
 nytimesbooks	& Edu-Books & 59.1\%	& 1.9\%\\
\hline
 EconUS	& Business & 74.4\%	& 	28.0\% \\
\hline
 EconWhichMBA	& Education & 37.6\% & 3.3\% \\ \cline{2-4}
		& Business  & 32.1\% & 	28.0\% \\
\hline
 PostSports	& Sports & 88.5\%    & 	9.6\% \\
\hline
 PostHealthSci	&	Science & 34.5\%	& 	0.96\% \\ \cline{2-4}
 		&	 Health & 25.1\% 	& 	5.3\%	\\
\hline
 WaPoFood	& Food 		& 60.3\%	&	6.3\% \\
\hline
\end{tabular}
\caption{{\bf Examples of topic-specific Twitter accounts of news organisations, along with the contribution of their topics of specialization in their production diet.}}
\label{tab:mass-media-subaccounts}
\end{table}

Interestingly, we find that each of the three news organizations has multiple accounts on Twitter. 
These include one primary account (@nytimes, @washingtonpost and @economist) 
and several {\it topic-specific accounts} (e.g., @NYTSports, @EconSciTech, @PostHealthSci) each
of which specializes in posting news-stories on a particular topic.
Table~\ref{tab:mass-media-subaccounts} shows 
some of the topic-specific accounts of the three news organizations, along with the fraction
of their production diet that is on the topic of specialization. 
It is evident that the topic-specific accounts produce a much larger fraction of their diet
on their specific topics of specialization, as compared to the mass media diet of the same news organization.

While the topic-specific accounts of the news organizations have thousands to hundreds of thousands of followers,
a much larger number of users subscribe to the primary accounts.
For instance, the primary account @nytimes has 15 million followers, while the topic-specific accounts 
@NYTSports and @nytimesbusiness have 51K and 567K followers respectively.
Since most social media users consume the diet produced by the primary account, we compare
the social media diet produced by the primary account with the mass media diet of the same news organization.

Figure~\ref{fig:media-compare} compares 
the information diets produced by the three news organizations over mass media, with
those produced by their primary Twitter accounts over social media. 
We find two main differences between the mass media and social media diets of the same news organization.
First, the primary accounts of the news organizations in social media tend to publish
less content (as compared to the corresponding mass media diets) on those topics
for which there exist topic-specific accounts.
For instance, for both NYTimes and Washington Post, topics such as sports and food
are covered much lesser in the social media diets than in the corresponding mass media diets.
Additionally, both the primary and the topic-specific social media accounts of the news organizations 
tend to be more specialized in their production by focusing on fewer topics, as compared to their mass media diets.
For example, while the mass media diet of Economist focuses on both business and politics,
the social media diet of @economist focuses solely on business and publishes
far lesser content on politics.

In summary, there is an {\it unbundling of content on social media} by the news organizations
through multiple accounts each specializing on a particular topic.
This unbundling would enable users in social media to get focused
information on their topics of interest by subscribing to the topic-specific accounts.
However, the users who subscribe to only the primary account of the news organizations
might not be aware that they are receiving a different information diet
as compared to that of the mass media versions.

\subsection{Popular social media accounts vs. mass media}

Next, we study whether our observations about the specialized production of
the social media accounts of news organizations generalizes to
other popular user-accounts in Twitter.
There are several ways to identify popular / influential accounts in Twitter, such as by the number of followers,
or by the number of times one is retweeted.
In this study, we consider {\it verified users} as examples of popular user-accounts on Twitter.
Out of all the verified users on Twitter who declared their language as English, and
were not news organizations, we randomly selected a set of 500 verified users. 
We collected the tweets posted by them during December 2014, and computed
the information diet posted by these users by the methodology presented earlier.

\begin{figure}[tb]
\center
\includegraphics[height=0.9\columnwidth,width=3.5cm,angle=-90]{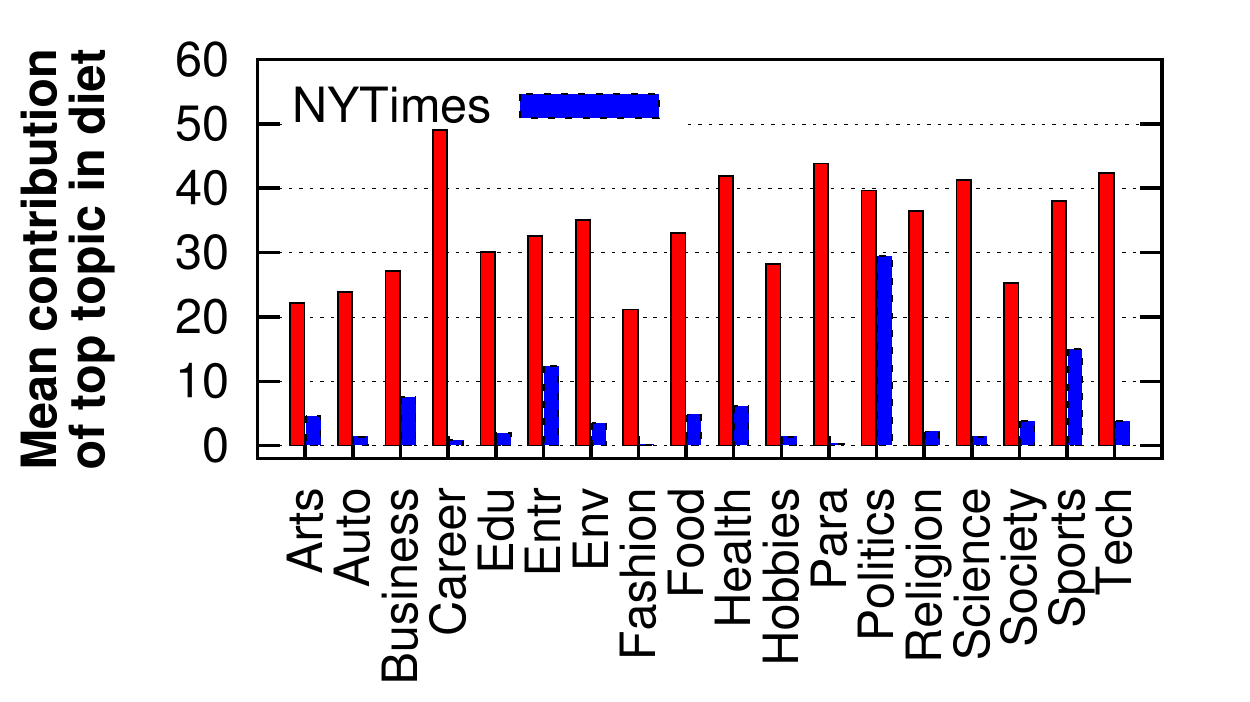}
\caption{{\bf Mean contribution of the top topic (on which a user posts the highest fraction of her diet) for popular users who focus on different topics.}}
\label{fig:verified-production-topic-wise-mean-coverage}
\end{figure}

For studying the specialization in the produced diet of each user, 
we define the {\it top topic} for her as the topic on which she posts the
largest fraction of her diet.
For the group of users having a common top topic,
we compute the mean percentage contribution of their posted diet that is on their top topic.
Figure~\ref{fig:verified-production-topic-wise-mean-coverage} shows this  mean percentage
contribution for the group of users specializing on each topic.
As a baseline, we also show the contribution of each topic in the NYTimes mass media diet (which was
stated in Table~\ref{table:mass-media-diet}).
We find that the popular users, on average, post a significant fraction of
their diet (between 20\% and 50\%) on just their top topic. 
Further, users having different top topics are focused to different degrees --
for instance, popular users having career, health, paranormal, science and technology
as their top topic post more than 40\% of their diet on their top topic.
Anyone who subscribes to these popular sources of information on social media
will get a much higher fraction of content on the corresponding topic,
than what is obtained from a typical mass media source 
(as shown by the NYTimes baseline in Figure~\ref{fig:verified-production-topic-wise-mean-coverage}).

\begin{figure}[tb]
\center
\includegraphics[height=0.9\columnwidth,width=3.5cm,angle=-90]{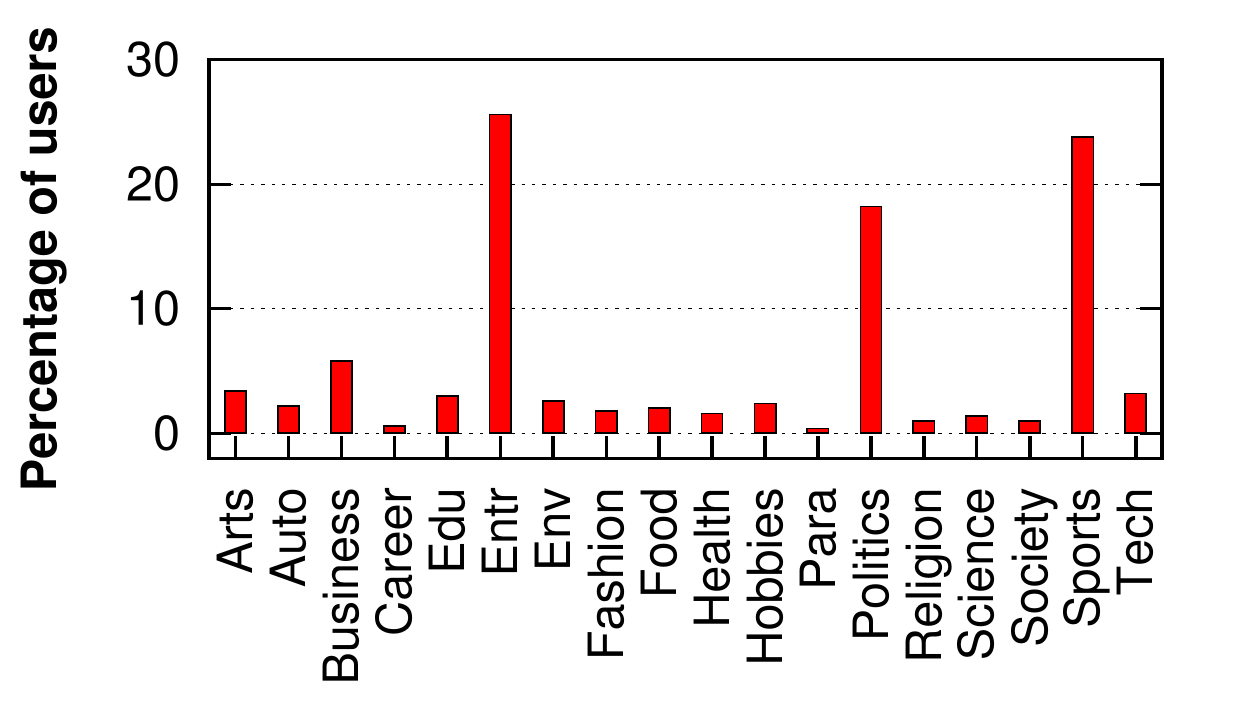}
\caption{{\bf Distribution of the 500 randomly selected verified users, according to the topic on which they produce the maximum fraction of their diet.}}
\label{fig:verified-users-topical-dist}
\end{figure}

Additionally, we looked at the distribution of the 500 randomly selected verified users across their top topics.
Figure~\ref{fig:verified-users-topical-dist} shows the distribution of these users 
according to their top topic. 
Most of the users have their top topic as one of the three topics -- {\it entertainment}, {\it sports} and {\it politics}. 
However, there are small fractions of popular users focusing their diets on all the other topics as well.
These observations agree with recent findings~\cite{deep-twitter-diving} that though Twitter is primarily thought to be
associated with few popular topics such as entertainment, sports, and politics, 
there are popular accounts who are experts on a wide variety of topics.

These observations imply that, similar to mass media, there are sources
of information on a wide variety of topics in the Twitter social media. 
However, since every source produces a diet that is specialized on just a few topics, 
the consumers of information in social media need to be careful in deciding whom they
subscribe to, especially if they desire to get a topically balanced information diet.

\subsection{Random sampling of social vs. mass media posts}

\begin{figure}[tb]
\center
\includegraphics[height=0.9\columnwidth,width=4cm,angle=-90]{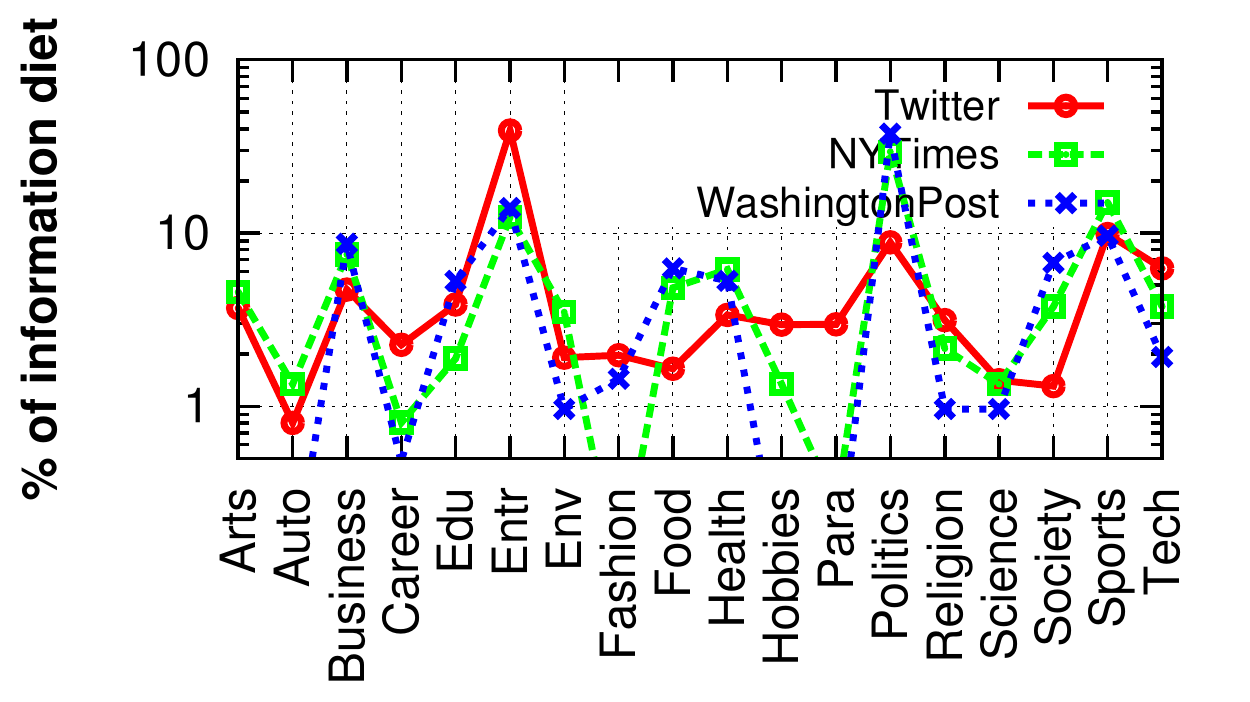}
\caption{{\bf Comparing the information diet of the Twitter 1\% random-sample with the mass media diet of news organizations (NYTimes and Washington Post).}}
\label{fig:twitter-massmedia-compare}
\end{figure}

Till now, we have shown that the individual sources of information in social media (popular user-accounts as well as accounts of news organizations) 
produce diets that are very focused on specific topics.
Now we shift the focus to the overall information being produced on the two media.
We use the Twitter 1\% random sample (for the month of December 2014) to represent the overall information 
being produced on social media, and compare the information diet of the Twitter random sample
with the mass media diets of NYTimes and Washington Post in Figure~\ref{fig:twitter-massmedia-compare}. 

We observe that the diets from both social media and mass media are skewed, but towards different topics. 
Though both diets have entertainment, politics, sports and business amongst the top topics, 
the Twitter social media diet is more heavily biased towards entertainment (39\%), while the mass media diets focus more on politics (30\%). 
Further, some topics are over-represented in the social media diet as compared to mass media diet, 
such as technology, hobbies-tourism, paranormal, and career.
On the other hand, topics such as food, health, and society are covered more in mass media than in social media,
which is probably because these topics are of general interest 
to many people in the off-line world. 
Whereas, topics such as entertainment and technology
are more dynamic, with new information being generated regularly, leading to them being covered 
more in a real-time information dissemination medium like Twitter.

\section{Consumption: Diets of Social Media Users}  \label{sec:consumption}

Unlike in mass media where everyone consumes the same broadcast information, every user on social media 
shapes her own personalized channel of consumption by subscribing to other users. 
In this section, we study how the users are consuming information 
in social media, as compared to the consumption via mass media.

For this analysis, we selected 500 users randomly from the Twitter userid space (i.e.,
the user-ids were randomly selected from the range 1 through the id assigned to a newly created account),
with the constraint that the selected users follow at least 20 other users (to ensure that the selected users 
have a meaningful consumption behaviour to study). 
We then computed the consumed information diet for each user, considering
the tweets that a user received from her followings (i.e., via word-of-mouth) during the month of December 2014.\footnote{We consider all tweets received by a user to compute her consumption diet, in the absence of data about what she actually reads.}

\begin{figure}[tb]
\center
\includegraphics[height=0.9\columnwidth,width=3.5cm,angle=-90]{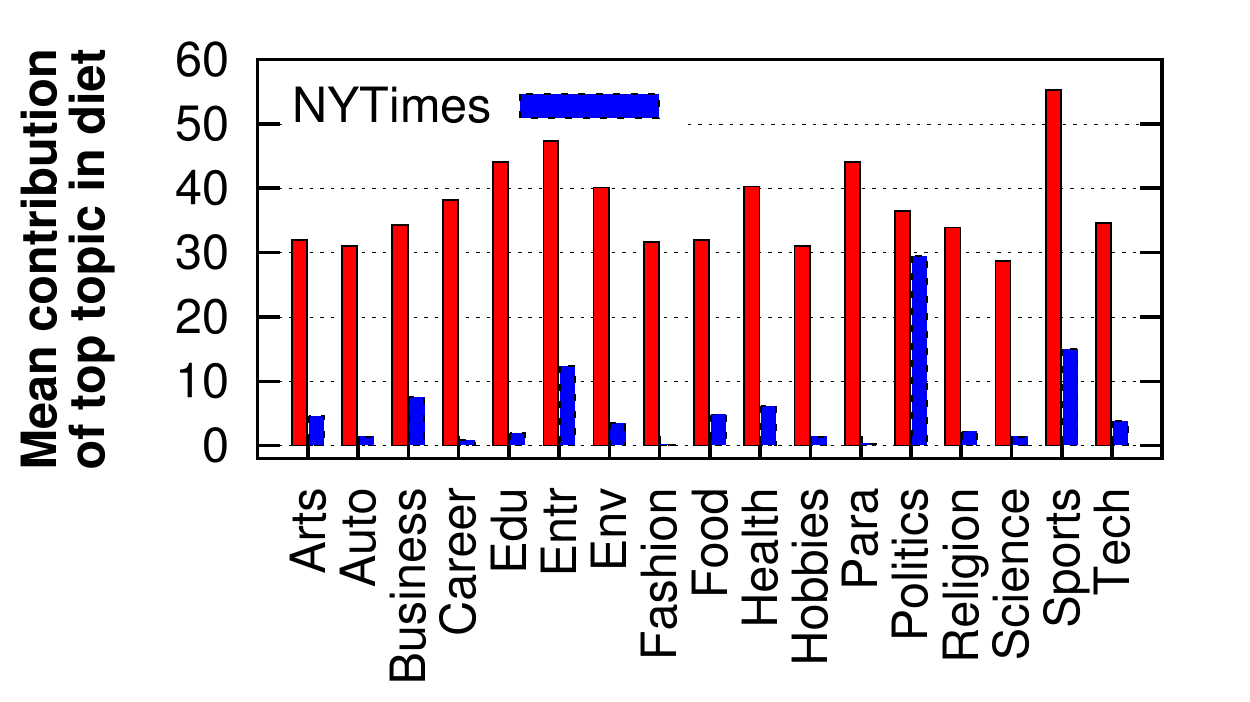}
\caption{{\bf Mean contribution of the top topic in the consumption diets of random users grouped according to their top topic of consumption.}}
\label{fig:random-consumption-topic-wise-mean-coverage-top-1-topic}
\end{figure}

\begin{figure}[tb]
\center
\includegraphics[height=0.9\columnwidth,width=3.5cm,angle=-90]{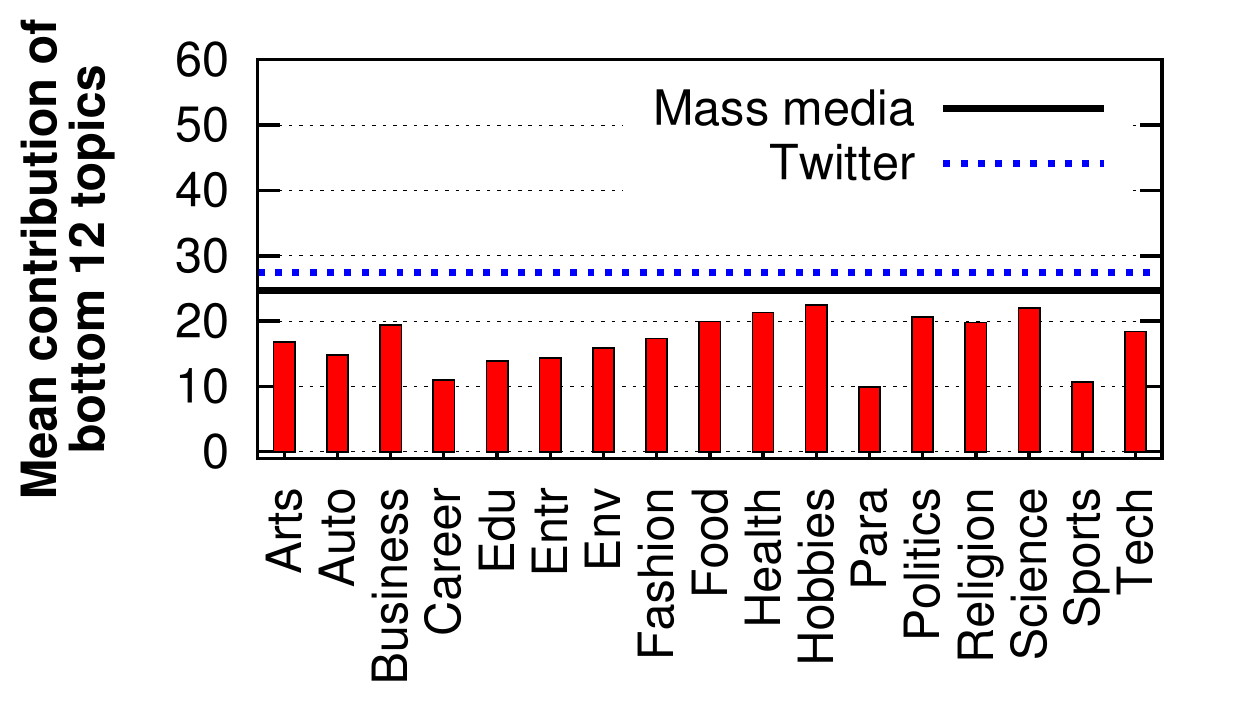}
\caption{{\bf Mean contribution of the bottom 12 least dominant topics in the consumption diets of random users grouped according to their top topic of consumption.}}
\label{fig:random-consumption-topic-wise-mean-coverage-bottom-12-topics}
\end{figure}

Similar to the previous section, we define the top topic
for a user as the topic on which she consumes the largest fraction of her diet.
For the group of users having a common top topic of consumption, 
Figure~\ref{fig:random-consumption-topic-wise-mean-coverage-top-1-topic} plots the mean contribution
of the top topic in the consumption diet of these users.\footnote{In our set of 500 randomly selected users, we did not find any user whose
 top topic of consumption was `society'; hence we will not consider this topic further in this section.} 
As a baseline, the figure also shows the contribution of each topic in the NYTimes
mass media diet.
Across almost all topics, the consumers are very focused on their top topic,
and on average, consume 30\% or more of their diet on that topic.
Moreover, when we compute the contribution of up to top two topics, 
we observe that 80\% of the users consume more than half of their diet 
on only these one or two topics.
These observations imply that users in social media consume a much larger fraction of their information diet
on their primary topic(s) of interest, as compared to what they would consume on the same topics
from a typical mass media source (as shown by the NYTimes mass media baseline).

Additionally, Figure~\ref{fig:random-consumption-topic-wise-mean-coverage-bottom-12-topics}
depicts the mean contribution of the {\it bottom 12 topics} on which the users consume the least information, for the same groups of users.
We find that the `tail topics' account for an inordinately low fraction of their consumed diet.
Across all topics, the mean tail topics contribution for users focusing on a particular topic 
is even lower than the contribution of the bottom 12 topics in the 
NYTimes mass media diet (24\%) and the Twitter random sample diet (27\%).

\begin{figure}[tb]
\center
\includegraphics[height=0.9\columnwidth,width=3.5cm,angle=-90]{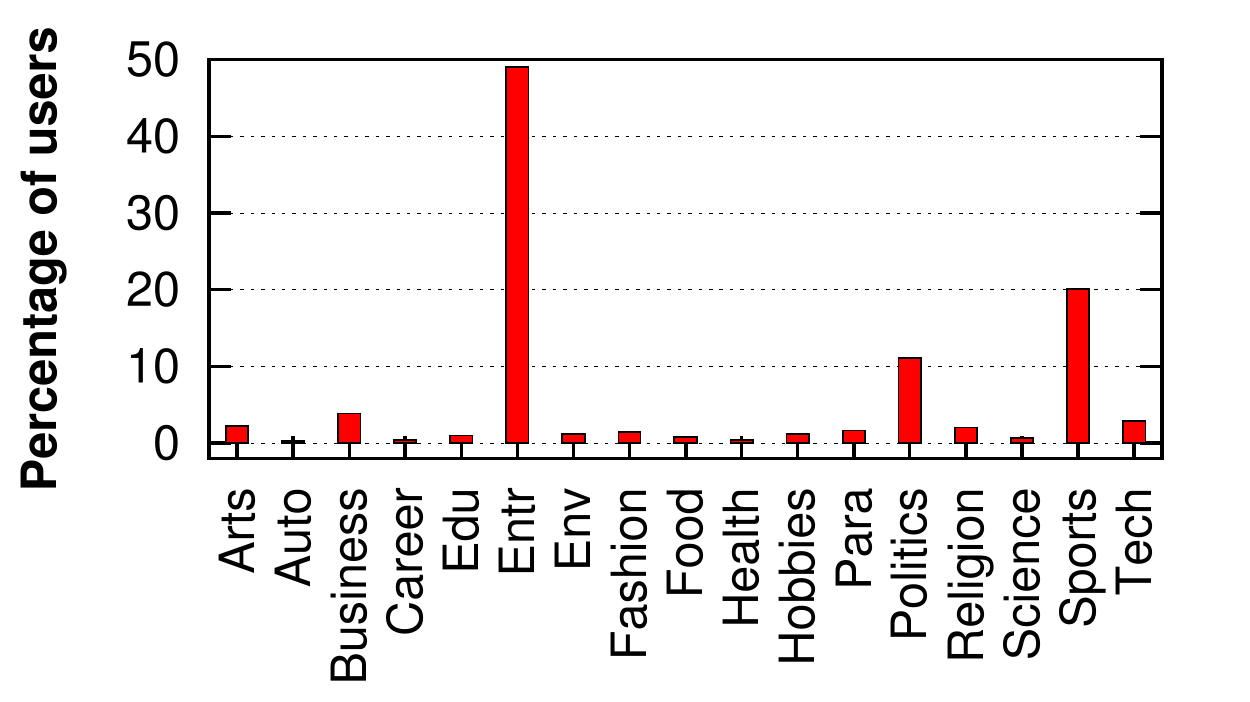}
\caption{{\bf Distribution of the 500 randomly selected users, according to the topic on which they consume the maximum fraction of their diet.}}
\label{fig:random-users-topical-dist}
\end{figure}

Finally, Figure~\ref{fig:random-users-topical-dist} plots the distribution of the 500 randomly selected users according to the top topic of consumption.
We find that the users' consumptions are very unevenly spread amongst the different topics --
as much as half the user population consumes most information on the topic entertainment, while
a sizeable fraction focuses on sports and politics.
When we compare this distribution to the production distribution of popular users in Figure~\ref{fig:verified-users-topical-dist}, 
we observe that consumption behaviours are even more skewed across topics than the production.

Thus we observe that 
users are extremely selective in the information they consume via social media,
with a huge bias towards one or two topics of their interest;
moreover, this bias comes at the cost of the tail topics. 
In future, as users rely more and more on social media like Twitter to consume information,
their diets may get progressively more skewed towards the one or two topics of their interest.
Users who wish to have a more balanced consumption in social media need to be careful about the 
sources to which they subscribe.
Alternatively, the biases in the consumption diets of users can potentially be mitigated
by the information supplied to them by recommender systems deployed in the social media sites;
in the next section, we investigate the role of recommender systems in shaping the
diets that social media users consume.

\section{Recommendations: Personalisation of Diets}  \label{sec:recessys}

All popular social media systems, including Twitter,
deploy recommendation systems to enable users discover content that would be interesting to them.
These recommendations expose the users to additional information beyond the information
which they get via word-of-mouth over their social follow-links. 
The recommendation systems currently deployed on most social media largely depend on the (2 hop) social neighborhood 
of a target user for finding interesting content to recommend to the user~\cite{facebook-reco,twitter-discover-tab-improved-personalization,twitter-reco}. 
Hence, such systems are also referred to as {\it social recommendation systems}. 

In the previous section, we saw that the consumed diets of most users are focused on just one or two topics 
of their interest. 
In this section, we study the impact of tweet recommendations on the information 
that users are exposed to, i.e., whether the recommendations exacerbate or mitigate 
the topical biases in the consumed diets.

\subsection{Data collection \& methodology}

On Twitter, the recommendations provided to a certain user are visible only to her 
and cannot be crawled publicly. Hence we adopt the methodology of creating 
{\it test accounts} on Twitter which {\it mimic} the followings of real users, 
i.e., the test accounts have the same network neighborhood as the mimicked real users. 
We randomly selected 15 real users with their number of followings varying between 10 and 1000
(to ensure that these users have social neighborhoods of different sizes),
and created test accounts mimicking 
these users.\footnote{These are passive test accounts which do {\it not} perform any activity such as
tweeting or favoriting, etc. They only gather the recommendations given to them by Twitter.
Even though the creation of such test accounts results in some users gaining an extra follower,
we believe that this has negligible effect on a large social network like Twitter.}
We refer to these test accounts as u1, u2, ..., u15.

The recommendations given in Twitter are dynamic, and are updated in real-time~\cite{twitter-reco}.
Hence, for each test account, we gathered a snapshot of the recommendations every 30 minutes, 
for a week in December 2014. On an average, each user received 708 recommended tweets 
in each gathered snapshot. 
Since these are too many for any user to view practically, we considered only the top 10 recommended 
tweets per snapshot.\footnote{We verified that the insights presented later in the section
hold even if we consider all recommended tweets (instead of the top 10).} 
We also collected the tweets received by each test account from all her followings, during the same period in December 2014.

For each of the 15 test accounts, we construct 3 information diets: 
(i)~{\it consumed diet}: the tweets it receives directly from the users it is following, 
(ii)~{\it recommended diet}: the top tweets recommended to it, and 
(iii)~{\it combined diet}: assuming that the user pays equal attention to the consumed and the recommended diets, this is constructed by 
considering the average contribution from consumed and recommended diets for each topic. 

\subsection{Recommended diets vs. Consumed diets}

We first investigate {\it whether the recommendations are personalized for each user}, 
i.e., whether different users get different recommended diets. 
Table~\ref{table:recommended-topicwise-contrib} states the
variation (range) in the percentage contribution of some of the topics 
in the recommended diets given to the 15 test accounts.
It is evident that different accounts are being recommended 
different diets, with varying contributions of topics.

\begin{table}
\small
\center
\begin{tabular}{|l|c||l|c|}
\hline
{\bf Topic} & {\bf Range (\%)} & {\bf Topic} & {\bf Range (\%)} \\ 
\hline
\hline
Automotive & 0.59 -- 10.83 & Business & 2.01 -- 18.01 \\ 
\hline
{\bf Entertainment} & {\bf 5.14 -- 40.36} & Environment & 1.27 -- 6.11 \\ 
\hline
Food & 0.49 -- 4.32 & Health & 0.79 -- 5.45 \\
\hline
{\bf Politics} & {\bf 9.03 -- 33.34} & Religion & 1.76 -- 6.81 \\
\hline
Science & 3.57 -- 13.05 & {\bf Sports} & {\bf 6.14 -- 46.97} \\
\hline
\end{tabular}
\caption{{\bf Range of contributions of different topics in the recommended diets given to the test accounts.}}
\label{table:recommended-topicwise-contrib}
\end{table}


Next, we examine {\it the extent to which the recommendations
given to a certain user match the consumed diet of the user}. 
In other words, assuming that the top topics in the consumed diet 
reflect the topical interests of the user, does the recommended diet contain
more or less of the same topics?

To quantify how well the recommended diet matches the consumed diet of a user,
we use the standard measure of KL-divergence 
(KLdiv in short) of the recommended diet from the consumed diet. 
The smaller the value of KLdiv, the closer the two diets are. 
We observe that the KLdiv values for the 15 test accounts vary in the range of 0.043 to 0.893, 
with 5 accounts having KLdiv values below 0.2, and 3 having values above 0.4.
This variation in the KLdiv values suggests that the recommendations 
match the consumed diets to different extents for different users. 

\begin{figure}[tb]
\center
\subfloat[{\bf u15 (min KLdiv: 0.043)}]{\includegraphics[height=0.47\columnwidth,width=3.8cm,angle=-90]{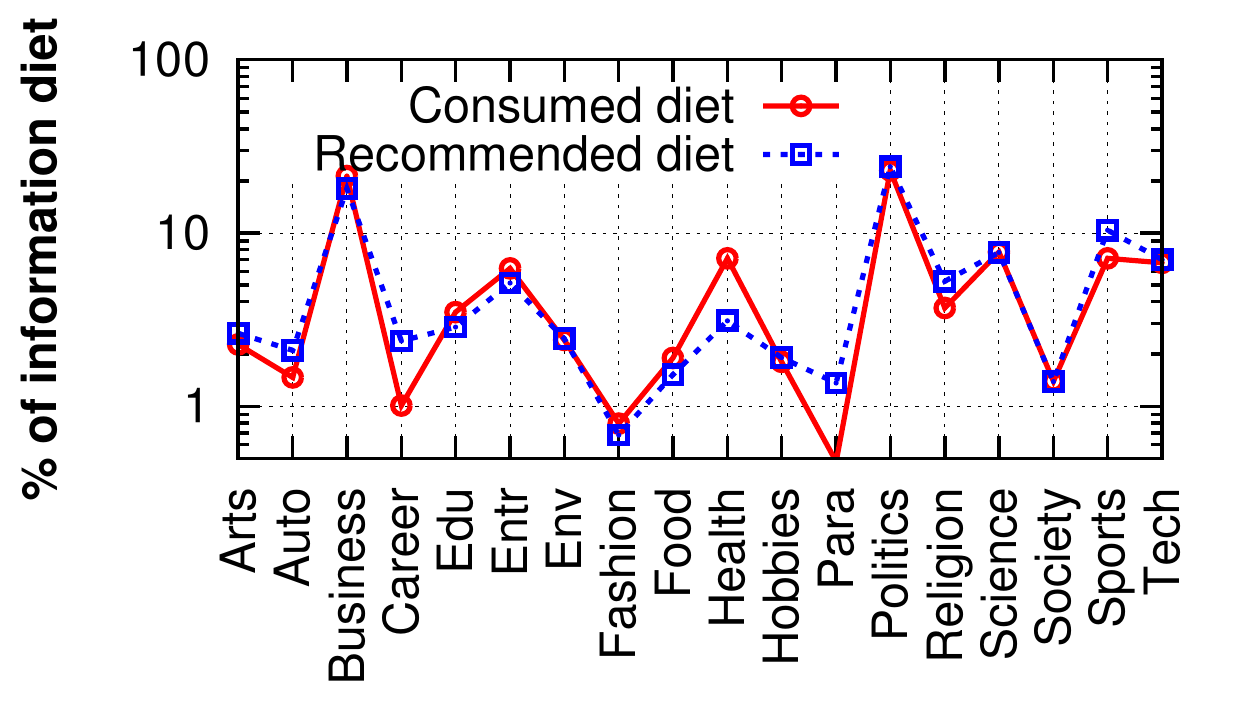}}
\hfil
\subfloat[{\bf u4 (max KLdiv: 0.893)}]{\includegraphics[height=0.47\columnwidth,width=3.8cm,angle=-90]{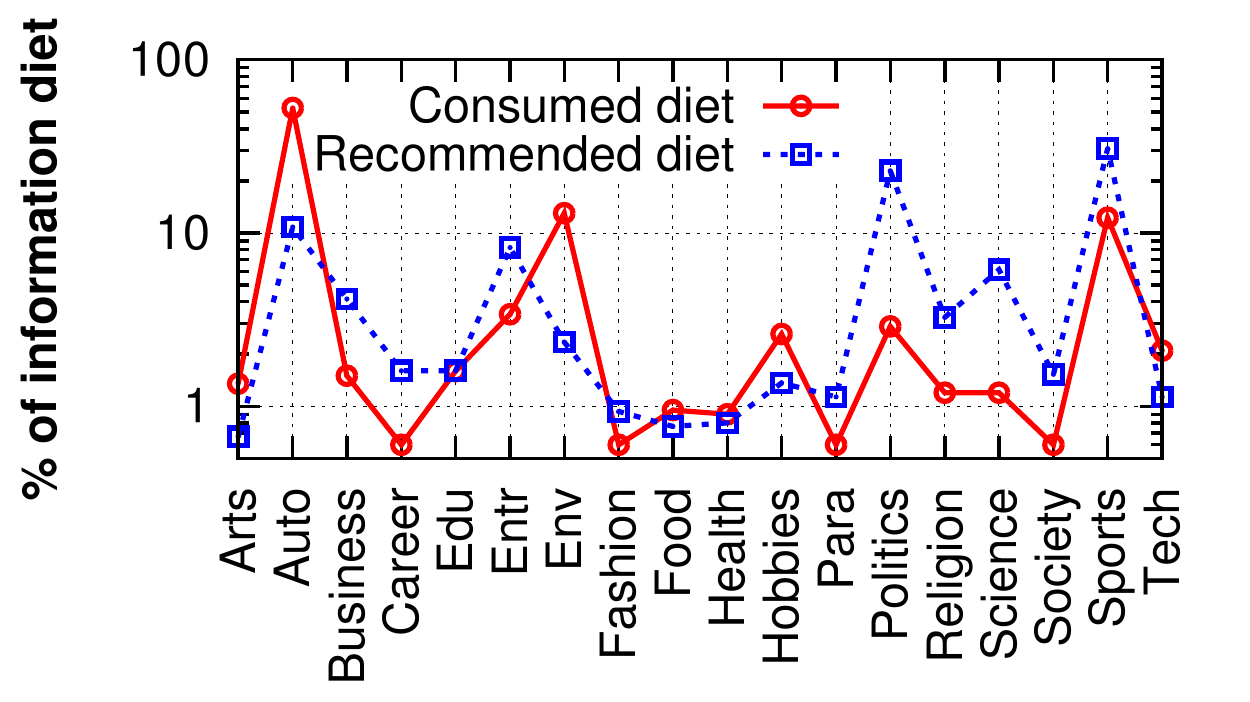}}
\caption{{\bf Comparing the consumed diet and recommended diet of two test accounts -- (i)~the one with the minimum KL divergence, and (ii)~the one with the maximum KL divergence of the recommended diet from the consumed diet.}}
\label{fig:consum-reco-1}
\end{figure}

Figure~\ref{fig:consum-reco-1} shows the topical compositions of the consumed and recommended diets 
for two test accounts --
(i)~u15 which has the minimum KLdiv, and (ii)~u4 which has the maximum KLdiv of the recommended diets from
their consumed diets. 
It can be seen that the recommended diet of u15 largely matches the consumed diet, 
while for u4 there is greater mismatch between the two diets. 
For instance, though u4 consumes a lot of information on the topics automobile and environment, 
its recommended diet has much lower fraction of these topics.
On the other hand, the recommended diet for u4 has higher fractions of 
politics, religion, and science, topics which are not
that significant in its consumed diet. 

These observations suggest that the recommended diet that
a user will get, does not always match her consumed diet.
We also observe cases where two accounts are consuming approximately the same amount of information 
on a particular topic, but they receive very different amounts of information 
on this topic in their recommended diets. 
These differences may be driven by the fact that different users have different
social neighborhoods, and the social recommendations given by Twitter
are derived from what information is popular in the social neighborhood of the user~\cite{twitter-reco}.

\begin{figure}[tb]
\center
\subfloat[{\bf Top 3 topics }]{\includegraphics[height=0.47\columnwidth,width=3.5cm,angle=-90]{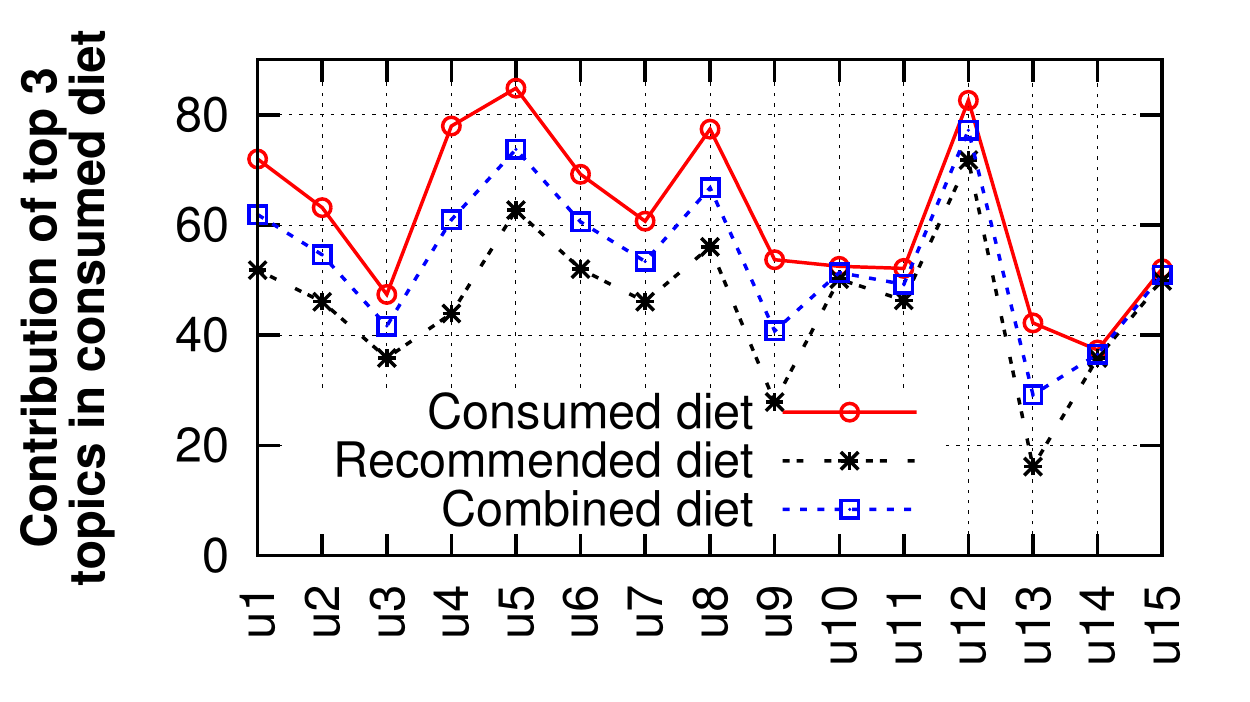}}
\hfil
\subfloat[{\bf Bottom 12 topics}]{\includegraphics[height=0.47\columnwidth,width=3.5cm,angle=-90]{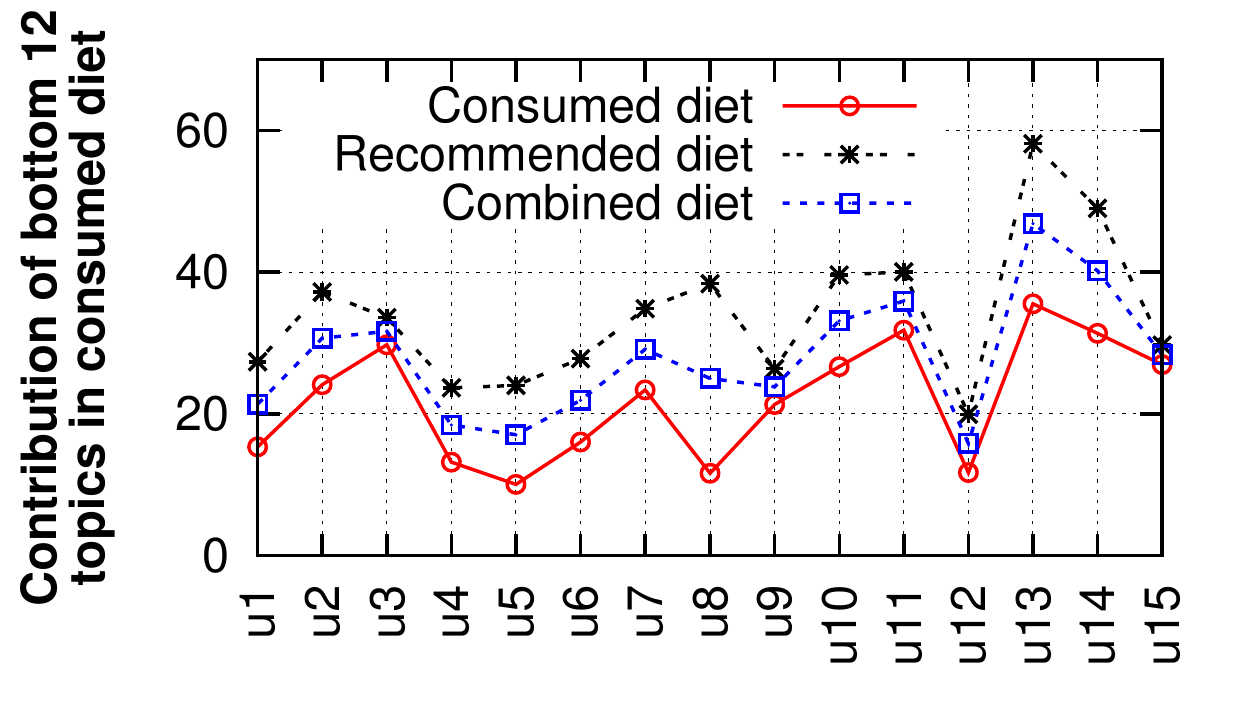}}
\caption{{\bf Contribution of the (i)~top 3 consumed topics and (ii)~bottom 12 consumed topics in the consumed, recommended and combined diets for the test accounts.}}
\label{fig:clones-consumption-rec-combined-diets}
\end{figure}

The effect of the social neighborhood can also be observed from Table~\ref{table:recommended-topicwise-contrib}
where it is seen that popular topics like entertainment, politics and sports 
are being recommended to everyone irrespective of whether they are interested in these topics. 
Every account is getting recommended at least 5\%, 9\% and 6\% in entertainment, politics and sports respectively, 
which is significantly higher than for other topics. 
On an average, every test account receives up to 17\%, 19\% and 17\% on entertainment, politics and sports respectively. 
As observed in earlier sections, there are a large number of users tweeting about these topics of general interest
(see Figure~\ref{fig:verified-users-topical-dist}), and hence everyone's neighborhood 
is likely to contain significant discussions on these topics, which get included into the social recommendations.

\subsection{Comparing with mass media diet}

Finally, we address the question {\it whether the recommendations mitigate or exacerbate the biases in the users' consumed diets}.
For this, we consider the top 3 topics in the consumed diet of an account (i.e., the 3 topics on which
the account consumes most information from its followings), and 
measure the contribution of these 3 topics in the consumed, recommended and combined diets of the user.
These are plotted for the 15 test accounts in Fig.~\ref{fig:clones-consumption-rec-combined-diets}(a).
Similarly, the Fig.~\ref{fig:clones-consumption-rec-combined-diets}(b) shows the contribution of 
the bottom 12 topics in the consumed diet of an account in the three diets.

Interestingly, we observe that the top 3 consumed topics account for a significantly smaller share in the recommended diets of the users, as compared
to the consumed diets. As a result, the combined diets of the users also contain a lesser contribution from these three topics, as compared to the consumed diets.
Again, the contribution of the bottom 12 topics is higher for the recommended and combined diets,
as compared to the consumed diets of the users. 
Thus, the recommendations tend to even out the imbalances in the consumed diets of the users, 
by including information from the lower ranking topics in user's consumed diets. 
Hence, social recommendations are reducing the gap between the information that different users are exposed to 
by mitigating the biases in the user's diets. 
To quantify this mitigation, we computed the KL-divergence between a user's (i)~consumed and (ii)~combined diets, from the 
baseline of the NYTimes mass media diet. 
We found that, for each of the accounts, the divergence from the baseline is lesser for the combined diet 
than for the consumed diet, showing that the social recommendations 
are actually having an equalizing effect across the users
(and driving the combined diets towards the baseline).

Thus, we find that social recommendations mitigate the imbalances
in the users' consumed diets, bringing in more heterogeneity into what the users are being exposed to.

\section{Concluding Discussion}  \label{sec:conclu}

In this work, we introduced the concept of {\it information diet} which is the 
topical composition of the information that is consumed or produced by a user. 
We proposed a novel methodology for quantifying information diets,
by inferring the topics of tweets and keywords in the Twitter social media. 
Our findings show that 
(i)~individual information sources (user-accounts) on social media 
produce information that is very focused on a few topics, 
(ii)~most users consume information primarily on one or two topics, 
and are often not careful about shaping a balanced diet for themselves, 
and 
(iii)~social recommendations somewhat mitigate the imbalances in the users' consumed diets by adding some topical diversity.

We envisage that this work will not only create awareness among social media users 
about potential imbalances in their information diets, 
but will also have implications for the designers of future information discovery, 
curation and recommendation systems for social media.
For instance, we found that social recommender systems are bringing in more heterogeneity into what the users are being exposed to. While this is good for broadening the horizons for the users, topic-specific recommendations 
might be necessary to provide information focused on the users' interests.
Studying the information diets provided by different types of recommender systems,
and their impact on the information that a user is exposed to, is an interesting direction
to pursue in the future.

~\\
\noindent {\bf Acknowledgements:}
The authors thank the anonymous reviewers whose suggestions helped to improve the paper.
S. Ghosh was supported by a post-doctoral fellowship from 
the Alexander von Humboldt Foundation.


\small{

}

\end{document}